\documentclass[prd,showpacs,superscriptaddress,eqsecnum,amsmath,amssymb,nofootinbib]{revtex4-2}

\usepackage{graphicx}
\usepackage{dcolumn}
\usepackage{bm}

\newcommand{\bea}{\begin{eqnarray}}
\newcommand{\eea}{\end{eqnarray}}

\usepackage{color}

\newcommand{\hide}[1]{}

\begin{document}

\title{Quantum Improved Regular Kerr Black Holes}

\author{Chiang-Mei Chen} \email{cmchen@phy.ncu.edu.tw}
\affiliation{Department of Physics, National Central University, Zhongli, Taoyuan 320317, Taiwan}
\affiliation{Center for High Energy and High Field Physics (CHiP), National Central University,
 Zhongli, Taoyuan 320317, Taiwan}

\author{Yi Chen} \email{yi592401@gmail.com}
\affiliation{Department of Physics, National Central University, Zhongli, Taoyuan 320317, Taiwan}

\author{Akihiro Ishibashi} \email{akihiro@phys.kindai.ac.jp}
\affiliation{Department of Physics, Kindai University, Higashi-Osaka, Osaka 577-8502, Japan}
\affiliation{Research Institute for Science and Technology, Kindai University, Higashi-Osaka, Osaka 577-8502, Japan}

\author{Nobuyoshi Ohta} \email{ohtan.gm@gmail.com}
\affiliation{Department of Physics, National Central University, Zhongli, Taoyuan 320317, Taiwan}
\affiliation{Research Institute for Science and Technology, Kindai University, Higashi-Osaka, Osaka 577-8502, Japan}


\begin{abstract}
We study the quantum improvement of Kerr black holes with mass-dependent scale identifications in asymptotically safe gravity. We find that a physically sensible identification can only be a function of $M r$ and the area $A = 4 \pi (r^2 + a^2)$ where $M$ is the mass of the black hole and $a$ is an angular momentum parameter. We then discuss various properties of the rotating quantum black holes for a simple choice of the identification. We show that the resulting regular rotating black holes have the following nice properties:
(i) admitting a consistent black hole thermodynamics at the horizon,
(ii) resolving the ring singularity,
(iii) partially eliminating closed time-like curves present in the classical Kerr black holes.
\end{abstract}


\maketitle

\section{Introduction}

The theory of general relativity has a universal and robust nature that gravitation produces singularities; the examples include the singularity in black holes and the big bang singularity in cosmology~\cite{Penrose:1964wq, Hawking:1967ju, Hawking:1970zqf}. The existence of singularity invalidates the applicability of general relativity. Therefore, many possible ways were proposed to avoid physical troubles caused by the singularities. The cosmic censorship conjecture presumes that a physical world should not have a causal connection with singularities~\cite{Penrose:1969pc}. The singularities in black holes should also never be ``naked''. Namely, they should be covered by a horizon. Another serious problem is that closed time-like curves (CTCs) can exist in solutions of general relativity. The existence of CTCs implies a theoretical possibility of traveling backward in time and then causes problems in causality. Indeed, it has been shown that CTCs do exist in Kerr black holes~\cite{Carter:1968rr}. The chronology protection conjecture was proposed that theory beyond general relativity, such as quantum gravity, should prevent time travel~\cite{Hawking:1991nk}. The cosmic censorship and chronology protection are proposals in classical theory. Alternatively the existence of likely unphysical singularity and CTCs could be just a consequence of incompleteness of general relativity which is a classical theory of gravity, and a complete theory of quantum gravity may resolve these problems. It would be desirable to study the possible resolutions of singularity and CTCs in the asymptotically safe gravity.

Several regular black holes have been proposed, for the review see~\cite{Torres:2022twv, Lan:2023cvz}, including Hayward~\cite{Hayward:2005gi} and Bardeen~\cite{Bardeen:1968} black holes. However, the focus in the previous proposals was mainly on the singularity resolution, but the preservation of black hole thermodynamics was not considered.

Among various approaches, the asymptotic safety scenario is an interesting proposal to formulate a quantum gravity theory using functional renormalization group~\cite{Reuter:1996cp, Souma:1999at}. For reviews, see~\cite{perbook, rsbook}. The quantum improvement, at solution level, may be made by replacing the coupling ``constants'' in the classical solutions with running couplings as functions of energy scale. The essential issue in this approach is a suitable choice of identification of the energy scale with some length scale in the considered solutions. This is often referred to as the ``scale identification.'' Many proposals have been studied for various black hole solutions~\cite{Bonanno:2000ep, Reuter:2004nv, Bonanno:2006eu, Falls:2010he, Reuter:2010xb, Harst:2011zx, Falls:2012nd, Koch:2013owa, Litim:2013gga, Koch:2014cqa, Bonanno:2017zen, Pawlowski:2018swz, Adeifeoba:2018ydh, Platania:2020lqb, Ishibashi:2021kmf, Ruiz:2021qfp, Chen:2022xjk, BP, Eichhorn:2022bgu, Platania:2023srt, Boos}.

There is a criticism that the gravitational coupling may not depend on the energy scale~\cite{Donoghue} because there are various configurations in any physical scattering process, and so there is intrinsic ambiguity in how to define the energy scale that is involved, and the gravitational corrections have different signs from the QED corrections. It may be true that simple energy dependence may not be enough to capture the whole effects of quantum gravitational effects, but gravitational coupling does depend on the energy scale and we believe that at least some aspects may be revealed by incorporating it. In particular we do not consider scattering amplitudes but geometrical configuration where we expect that there is a unique energy scale on any point. So here we take the viewpoint that we can use some kind of scale identification and study the physical consequences.

Quantum effects of gravity are expected to play a significant role in the core region near the black hole singularity. In the asymptotically safe gravity, it turns out that quantum effects provide a ``repulsive'' force at the core of black hole which stabilizes the unstable phase of small black holes and gives the possibility of resolving the singularity~\cite{Bonanno:2000ep, Bonanno:2006eu}. However a problem in this approach is that we do not have any clear physical criterion how to make the identification of the energy scale and the length scale. In Ref.~\cite{Reuter:2010xb}, the energy $k$ is identified with the inverse radial length from the origin to the point under consideration with {\it a fixed angle} for the Kerr black holes. With this scale identification, it turned out that either there exists no entropy-like state function or the temperature cannot be in proportion to the surface gravity. For this reason, it has been proposed in~\cite{Reuter:2010xb} that the definitions of the temperature and entropy should be modified in such a way that the first law is satisfied in the small angular-momentum $J$ expansions. In this proposal, the authors of \cite{Reuter:2010xb} were able to give the solution only to the order of $J^2$ but no complete expression. 
However, it is unnatural to consider that the temperature is modified, since it should be defined through the geometric quantity of the surface gravity at the horizon.

In Ref.~\cite{Chen:2022xjk}, we have considered this problem for the Kerr-(A)dS black holes.
This is a nontrivial problem because in the presence of angular momentum, the entropy may be regarded as a function of both the black hole mass and the angular momentum parameter, and its variation is given by
\begin{equation} \label{cons}
dS = \partial_{M} S \, dM + \partial_a S \, da,
\end{equation}
where $M$ is the black holes mass and $a$ is the angular momentum parameter. Unless the consistency or integrability condition $[\partial_M, \partial_a] S = 0$ is satisfied, the entropy is not well defined. This is the origin of the problem discussed in Ref.~\cite{Reuter:2010xb}. In Ref.~\cite{Chen:2022xjk}, we have pointed out that in order to satisfy the consistency condition, the scale identification should be such that the gravitational coupling is a function of the horizon area, at least in the neighborhood of the horizon,
\begin{equation} \label{area}
A_h = 4 \pi (r_h^2 + a^2).
\end{equation}
With this, we have found that the first law is exactly satisfied. Since the first law of thermodynamics is the fundamental law of energy conservation, it must be valid not only semiclassically but also quantum mechanically. Note that this is a general consequence from the first law if we allow coordinate dependence of the gravitational constant, and hence independent of the asymptotic safety. It is natural to extend the identification away from the horizon as a function of the area $A = 4 \pi (r^2 + a^2)$ at the fixed radius $r$. Unfortunately it is clear that this kind of the gravitational coupling depending only on $A$ cannot resolve the ring singularity at $r = 0$ because it never vanishes or diverges to cancel the singularity whatever the choice of the identification. This problem arises because in~\cite{Chen:2022xjk} we considered only the mass-independent identification. In this paper, we generalize our previous analysis to include the dependence on the mass parameter, and show that a more general scale identification which preserves the thermodynamic properties is possible and the gravitational coupling can be a function of $M r_h$ as well as the area $A_h$ on the horizon. Indeed, a simple mass-dependent identification was already considered in Ref.~\cite{Reuter:2010xb} but the thermodynamics was a nontrivial issue there, as discussed above. Actually the former variable should also be a function of the area on the horizon. It is then natural to extend this away from the horizon, keeping the dependence on both variables. It is crucial to notice here that when we leave away from the horizon, the area and $M r$ become independent. These are related only on the horizon, as we will show later. We then consider a simple physically sensible choice of a scale identification in the asymptotically safe gravity, which gives regular rotating black holes with the following nice properties:
(i) admitting a consistent black hole thermodynamics at horizon,
(ii) resolving the ring singularity,
(iii) partially eliminating CTCs.

We note that in a recent paper~\cite{Yang:2023cmv}, the problem is studied if known rotating regular black holes can satisfy the thermodynamics. Most of the works on rotating regular black holes consider the case when the mass or equivalently the gravitational constant is a function of coordinate, and the thermodynamics are not
satisfied. However our study~\cite{Chen:2022xjk} has shown that the running coupling or mass should be a function of horizon area in order to preserve the consistency of the first law of thermodynamics. So it is to be expected that the thermodynamics cannot be satisfied if one considers only the coordinate dependence.

When we make quantum improvement, there are several choices of coordinate systems such as Boyer-Linquist (BL) or Eddington-Finkelstein (EF) coordinates, which may give different solutions upon quantum improvement. At the classical level, they are related by a coordinate transformation and so they describe the same solution. However, if we include the angular dependence in the running gravitational constant, they are no longer the same solution. Here we show that the absence of the singularity on the horizon or some other places requires that such an angular dependence should not be included. Then the solutions are related by the coordinate transformation and we can study the problem in any coordinate, and find the above results in the BL coordinate system. Requiring the consistency with the thermodynamics, we find a quite general formula for the entropy
\begin{equation}
S = \int \frac{dA_h}{4 (G + M \partial_M G)},
\end{equation}
where $G$ is the gravitational constant dependent on $M r_h$ and area $A_h$.

As a bonus of our quantum improvement, we find that the CTC present in the Kerr solution may be also removed in the region where the quantum effects are significant. The precise discussions are given in Sec.~\ref{sec:4}.

This paper is organized as follows. In the next section, we briefly review the asymptotic safe quantum gravity. In Sec.~\ref{sec:3}, we start our study of scale identification for Kerr black holes with possible mass dependence. In Sec.~\ref{sec:3A}, after briefly reviewing some basic aspects of the classical Kerr geometry, we consider a scale identification as a function of the radial and angular coordinates in the BL coordinate, and show that the running gravitational coupling should not depend on the angle in order to avoid singularity on the horizon (this result itself was given before in~\cite{Held:2021vwd,Chen:2022xjk}). In Sec.~\ref{sec:3B}, we consider the solution in the EF coordinates where there is no curvature singularity even if we keep the angular dependence. It turns out that in this case, we encounter another problem of ``parallelly propagated curvature singularity,'' so the angular dependence is excluded. We then show in Sec.~\ref{sec:3C}, that the consistency with the thermodynamics requires that the gravitational coupling should be a function of $M r_h$ and the horizon area $A_h$, and give a general formula for the entropy. It is natural to extend the dependence away from the horizon, keeping the dependence on $M r$ and the area $A = 4 \pi (r^2 + a^2)$ at fixed radius $r$. We then discuss the properties of the quantum improved black holes in Sec.~\ref{sec:4}. First in Sec.~\ref{sec:4A}, we show that a simple choice of the scale identification can resolve the ring-singularity existed in the classical Kerr geometry. We also show the resolution of CTCs in Sec.~\ref{sec:4C}.

We summarize our results in Sec.~\ref{sec:5}. In appendix, we list the algebraically complete set of invariant curvature scalars for quantum improved Kerr black holes (Appendix~\ref{sec:AA}), and the analysis of quantum improved Kerr-(A)dS black holes (Appendix~\ref{sec:AB}). We briefly discuss the analytic continuation of null geodesics in our quantum improved Kerr black holes (Appendix~\ref{sec:4B}).

\section{Asymptotically Safe Gravity}\label{sec:2}

The asymptotically safe scenario for quantum generalization of the general relativity with cosmological constant, in the units $c = \hbar = 1, G = l_P^2 = 1/M_P^2$:
\begin{equation}
S = - \frac1{16 \pi G} \int d^4x \sqrt{-g} (R - 2 \Lambda),
\end{equation}
proposes an energy-scale $k$ dependent Newton constant $G(k)$ and cosmological constant $\Lambda(k)$~\cite{Reuter:1996cp}. Assuming that the cosmological constant is already at the fixed point and negligibly small,\footnote{
This assumption may be justified if we consider the wave function renormalization~\cite{KO}.}
the associated renormalization group equations lead to the solutions for the Newton coupling~\cite{Pawlowski:2018swz} 
\begin{equation}
G(k) = \frac{G_0}{1 + \omega G_0 k^2},
\label{Ncoupling}
\end{equation}
where $\omega$ is a constant of order 1 representing the quantum effects, and $G_0 = G(k=0)$. Note that the quantum effect becomes significant when the energy is closed to the Planck energy
\begin{equation} \label{eq_QE}
k^2 \sim (\omega G_0)^{-1} = M_P^2/\omega.
\end{equation}

In order to see what physical consequence this may have on the black hole geometry, we have to identify the energy scale $k$ with some length scale in the black hole geometry. This is an essential issue for constructing quantum improved solutions. A simple possibility is to assume that the energy scale is dependent on the radius and angular momentum parameter $a$ but independent of the mass parameter. The first law of black hole thermodynamics for the Kerr-(A)dS black holes then requires that the scale identification near the horizon should be a function of the horizon area $A_h$~\cite{Chen:2022xjk}. It is natural to generalize the scale identification away from the horizon by replacing the horizon radius $r_h$ [see below Eq.~(\ref{eq_KerrDS})] with the radial coordinate $r$, but then the energy scale determined by the area remains finite at the classical singularity ($r = 0, \theta = \pi/2$ for Kerr black holes) and cannot resolve the singularity.

In order to resolve the ring singularity for Kerr black holes, we have to extend our assumption. Note that it is also possible to include the dependence on the mass parameter in the scale identification. We will show in Sec.~\ref{sec:3} that there is a more general scale identification, corresponding to quantum improvement of the classical rotating Hayward black holes~\cite{Bambi:2013ufa}, which is consistent with black hole thermodynamics on the horizon and, when extended away from the horizon, can resolve the classical ring singularity. The results can be generalized to include a cosmological constant; we will discuss this in Appendix~\ref{sec:AB}.

\section{Running Coupling for Kerr Black Holes} \label{sec:3}

When we make scale identification, we find that the quantum improvements in different coordinate system give different solutions. In this section, we first discuss the identification in the BL coordinates and argue that the identification should not include the angular coordinate in order not to introduce the singularity on the horizon. It turns out that such a singularity does not appear if we use the EF coordinates. However in this case there appears another problem, so the identification should not have angular dependence either.

\subsection{Scale identification in the BL coordinates} \label{sec:3A}

Let us start with the Kerr black hole in the BL coordinates. The line-element of the classical solution is
\begin{equation} \label{eq_KerrBL}
ds^2 = - \frac{\Delta}{\Sigma} \left( dt - a \sin^2\theta d\varphi \right)^2 + \frac{\Sigma}{\Delta} dr^2 + \Sigma d\theta^2 + \frac{\sin^2\theta}{\Sigma} \left( a dt - (r^2 + a^2) d\varphi \right)^2,
\end{equation}
where
\begin{equation} \label{eq_KerrDS}
\Delta = r^2 - 2 G M r + a^2, \qquad \Sigma = r^2 + a^2 \cos^2\theta,
\end{equation}
and $a$ is the angular momentum parameter. In the semiclassical case, i.e. $G$ is a constant, the Hawking temperature and the entropy are given by
\begin{equation} \label{eq_TS}
T = \frac{\kappa}{2\pi} = \frac{r_h^2 - a^2}{4 \pi r_h (r_h^2 + a^2)}, \qquad
S = \frac{A_h}{4 G} = \frac{\pi (r_h^2 + a^2)}{G},
\end{equation}
where $r_h$ is the radius of the horizon determined by $\Delta(r_h) = 0$ and $A_h$ is the area of horizon~\eqref{area}.
The quantity $\kappa$ here is the surface gravity, which is defined as follows:
We choose $\Omega_H$ in the Killing vector $\chi = \partial_t + \Omega_H \partial_\varphi$ such that $\chi$ is null ($\chi_\mu \chi^\mu = 0$) at the horizon, and then the surface gravity $\kappa$ is given by the formula
\begin{equation} \label{eq_surfaceG}
\kappa^2 = - \frac{(\nabla_\mu \chi_\nu) (\nabla^\mu \chi^\nu)}{2}.
\end{equation}
Note the second expression in~\eqref{eq_TS} for the entropy is the standard Bekenstein-Hawking formula. These thermal quantities satisfy the first law of thermodynamics
\begin{equation} \label{eq_firstlaw}
dM = T dS + \Omega dJ,
\end{equation}
where angular momentum and angular velocity of black holes are
\begin{equation} \label{eq_JO}
J = M a, \qquad \Omega = \frac{a}{r_h^2 + a^2}.
\end{equation}

Note that the classical Kerr geometry (and the quantum improved one, too) admits two horizons, the (outer) event horizon with $r_h = G M + \sqrt{(G M)^2 - a^2}$ and the (inner) Cauchy horizon with $r_h = G M - \sqrt{(G M)^2 - a^2}$, and we can define the above thermodynamic quantities in the same manner for both the horizons. Throughout this paper, when we discuss the black hole thermodynamics, we are mainly concerned with the outer event horizon, while when we discuss the region near the singularity as below, we have in mind the area inside the Cauchy horizon.

The Kretschmann scalar is
\begin{equation} \label{KerrKretschmann}
\mathcal{K} = \frac{48 M^2 G^2 (r^6 - 15 a^2 \cos^2\theta r^4 + 15 a^4 \cos^4\theta r^2 - a^6 \cos^6\theta)} {(r^2 + a^2 \cos^2\theta)^6},
\end{equation}
which depends on the angle coordinate $\theta$. In the limit $r \to 0$, we have
\begin{equation}
\lim_{r \to 0} \mathcal{K} = \begin{cases} \dfrac{48 M^2 G^2}{r^6} \to \infty, & \theta = \pi/2 \\
 & \\
- \dfrac{48 M^2 G^2}{a^6 \cos^6\theta} < \infty , & \theta \ne \pi/2 \end{cases} \, .
\end{equation}
This indicates that there is a ring singularity at $r = 0, \, \theta = \pi/2, \, \varphi=\mbox{arbitrary}$, and the Kretschmann scalar is not a continuous function of $\theta$ on the disk $r = 0$ at the ring singularity.

In asymptotically safe gravity, the running coupling varies according to the energy scale. We assume that it can be a function of the coordinates of both $r$ and $\theta$ for Kerr black holes. However, we note some undesirable properties in the quantum improvements by an angle-dependent running coupling $G = G(k(r, \theta)) = G(r, \theta)$ in the metric~\eqref{eq_KerrBL}. Namely, the scalar curvature becomes
\begin{equation}
R = \frac{2 M}{\Sigma} \left[ r \partial_r^2 G(r, \theta) + 2 \partial_r G(r, \theta) - \frac{M r^2}{\Delta^2} \Bigl( \partial_\theta G(r, \theta) \Bigr)^2 \right],
\end{equation}
where $\Delta$ is the function in the Kerr solution~\eqref{eq_KerrDS}, and this vanishes on the horizon.
This indicates that a scalar curvature singularity occurs on the horizon when the coupling depends on the angle $\theta$. To avoid the singularity, the gravitational coupling must not depend on the angle $\theta$.

\subsection{Scale identification in EF Coordinates} \label{sec:3B}

We note that there are other coordinate systems for the Kerr black hole which are free from such a curvature singularity even if the gravitational coupling depends on the angle~\cite{Held:2021vwd}. For example, one can introduce the EF coordinates by
\begin{equation} \label{eq_EF}
du = dt + \frac{r^2 + a^2}{\Delta} dr, \qquad d\tilde\varphi = d\varphi - \frac{a}{\Delta} dr.
\end{equation}
The classical Kerr black holes then become
\begin{eqnarray}
ds^2 &=& - \frac{\Delta - a^2 \sin^2\theta}{\Sigma} du^2 + 2 du dr - \frac{2 a \sin^2\theta (r^2 + a^2 - \Delta)}{\Sigma} du d\tilde\varphi - 2 a \sin^2\theta dr d\tilde\varphi
\nonumber\\
&+& \Sigma d\theta^2 + \frac{(r^2 + a^2)^2 - \Delta a^2 \sin^2\theta}{\Sigma} \sin^2\theta d\tilde\varphi^2.
\end{eqnarray}
As pointed out in~\cite{Held:2021vwd}, unlike in BL coordinates, the angle-dependent running coupling $G = G(r, \theta)$ does not produce the curvature singularity at the horizon in this coordinate system. Indeed, in EF coordinates, all the algebraically complete set of four invariants~\cite{Torres:2016pgk} do not include any angle-derivative of Newton coupling, and thus they remain as in the expressions~\eqref{eq_KerrR}-\eqref{eq_KerrK}. However, the Ricci tensor does include  $\partial_\theta G$:
\begin{eqnarray}
&& R_{uu} = - \frac{M}{\Sigma^3} \Bigl[ r \Sigma \Delta \partial_r^2 G - 2 a^2 (r^2 \sin^2\theta - \Delta \cos^2\theta) \partial_r G + r \Sigma \partial_\theta^2 G + r \cot\theta (\Sigma + 4 a^2 \sin^2\theta)  \partial_\theta G \Bigr],
\nonumber\\
&& R_{ur} = \frac{M \left( r \Sigma \partial_r^2 G + 2 a^2 \cos^2\theta \partial_r G \right)}{\Sigma^2}, \qquad
R_{u\theta} = \frac{M \left[ r \Sigma \partial_r \partial_\theta G - (r^2 - a^2 \cos^2\theta) \partial_\theta G \right]}{\Sigma^2},
\nonumber\\
&& R_{u\tilde\varphi} = \frac{M a \sin^2\theta}{\Sigma^3} \Bigl[ r \Sigma \Delta \partial_r^2 G - 2 (r^4 + r^2 a^2 - a^2 \Delta \cos^2\theta) \partial_r G + r \Sigma \partial_\theta^2 G  + r \cot\theta (3 \Sigma + 4 a^2 \sin^2\theta) \partial_\theta G \Bigr],
\\
&& R_{r\tilde\varphi} = - a \sin^2\theta R_{ur}, \qquad R_{\theta\theta} = \frac{2 M r^2 \partial_r G}{\Sigma}, \qquad R_{\theta\tilde\varphi} = - a \sin^2\theta R_{u\theta},
\nonumber\\
&& R_{\tilde\varphi\tilde\varphi} = - \frac{M \sin^4\theta}{\Sigma^3} \left[ r a^2 \Sigma \Delta \partial_r^2 G - 2 \left( \frac{r^2 (r^2 + a^2)^2}{\sin^2\theta} - a^4 \cos^2\theta \Delta \right) \partial_r G + r a^2 \Sigma \partial_\theta^2 G + r a^2 \cot\theta (4 r^2 + 4 a^2 + \Sigma) \partial_\theta G \right].
\nonumber
\end{eqnarray}
Actually, the transformation~\eqref{eq_EF} is not a valid coordinate transformation if the running coupling depends on the angle $\theta$ (because then $\Delta$ also depends on $\theta$). Therefore the black holes are different when we apply the angle dependent identification to Kerr solution in BL and EF coordinates; they are not related by the coordinate transformation.

So in this coordinate system, it appears that the angle dependence may not be excluded. However, we encounter another problem here. Since our quantum improved black hole is stationary and axisymmetric, according to the black hole rigidity~\cite{Hawking:1971vc, Carter:1971zc}, the event horizon is expected to be a Killing horizon
to which both the two Killing vector fields $\partial_u$ and $\partial_{\tilde \varphi}$ are tangent,\footnote{
Note that we denote here the stationary and axial Killing vector fields in terms of the EF coordinates $(u, {\tilde \varphi})$.}
and hence should be determined by $\Delta (r_h) = 0$. However, the linear combination $\partial_u + \Omega_H \partial_\varphi$ with the horizon angular-velocity $\Omega_H := a/(r_h^2 + a^2)$, which is null on the horizon, fails to be a Killing vector as $\Omega_H$ is now a function of $\theta$ via $r_h$ with angular dependence through the horizon condition $\Delta(r_h) = 0$.
Even if one finds a null Killing vector field $\chi$ tangent to the horizon null generators, one may encounter the following physically unwanted problem. For simplicity, let us consider the Schwarzschild case, i.e., $a = 0$. In this case, $\chi = \partial_u$ itself is the horizon null Killing vector field and one can compute the Hawking temperature via~\eqref{eq_surfaceG} and obtain
\begin{eqnarray}
T = \frac{\kappa}{2 \pi} = \frac1{8 \pi G(r_h, \theta) M} - \frac{\partial_{r_h} G(r_h, \theta)}{4 \pi G(r_h, \theta)}.
\end{eqnarray}
This is no longer a constant over the horizon, and hence the zero-th law of black hole thermodynamics fails to hold. Furthermore, it is shown in general circumstances that if the surface gravity is not constant over the Killing horizon, then there must appear a ``parallelly propagated curvature singularity'' along horizon null generators~\cite{Racz:1992bp, Racz:1995nh}. The latter issue may be overlooked when one examines the behavior of the curvature scalars only [see Appendix~\ref{sec:AA}]. To remove these problems, the gravitational coupling has to be independent of the angle.

\subsection{Consistency in the BL coordinates} \label{sec:3C}

These observations lead us to consider that the scale-identified coupling should be a function of the coordinate $r$ and the physical parameters $M$ and $a$, namely $G = G(r, M, a)$ for rotating black holes in any coordinate system. Note that in this case,~\eqref{eq_EF} gives a coordinate transformation and we can use any coordinate system to discuss the quantum improvement. So from now on, we use the BL coordinate system to discuss the consistency with the thermodynamics.

The horizon radius $r_h$ of quantum improved Kerr black holes is determined by
\begin{equation} \label{hc}
\Delta(r_h) = r_h^2 + a^2 - 2 G(r_h, M, a) M r_h = 0.
\end{equation}
In general, it is difficult to find the explicit expression for $r_h$ in terms of $M$ and $a$, but we can have its first derivatives from the relation
\begin{eqnarray}
0 = d\Delta(r_h) &=& 2 r_h (\partial_M r_h \, dM + \partial_a r_h \, da) + 2 a \, da - 2 r_h G \, dM - 2 M r_h \partial_M G \, dM - 2 M r_h \partial_a G \, da
\nonumber\\
&-& 2 M (G + r_h \partial_{r_h} G) (\partial_M r_h \, dM + \partial_a r_h \, da),
\end{eqnarray}
leading to
\begin{equation} \label{eq_partialR}
\partial_M r_h = \frac{r_h (G + M \partial_M G)}{r_h - M G - M r_h \partial_{r_h} G}, \qquad
\partial_a r_h = - \frac{a - M r_h \partial_a G}{r_h - M G - M r_h \partial_{r_h} G}.
\end{equation}

In general, arbitrary identification does not guarantee to lead to a thermodynamically consistent system, and the consistency is an essential physical criterion for an acceptable scale identification. The mass-independent identification has been discussed in~\cite{Chen:2022xjk}, and it is shown that a consistent running coupling should be a function of area at horizon which tells us that the Newton coupling should be so; $G = G(r^2 + a^2)$. However, it is clear that such a Newton coupling cannot give a necessary condition $G(r = 0) = 0$ for resolving the singularity at $r = 0$~\cite{Torres:2022twv}. Here we examine the consistency for a more general identification depending on mass parameter as well.
Note that the dependence on the mass was also included in~\cite{Bonanno:2000ep} since the geodesic distance depends on it.

It is straightforward to compute the angular momentum, angular velocity and temperature (note that the numerator of $T$ equals to denominator of $\partial_M r_h$):
\begin{equation} \label{eq_T}
J = M a, \qquad \Omega = \frac{a}{r_h^2 + a^2}, \qquad
T = \frac{r_h - M G - M r_h \partial_{r_h} G}{2 \pi (r_h^2 + a^2)}.
\end{equation}
Then the thermodynamical first law $\delta M = T \delta S + \Omega \delta J$ leads to
\begin{equation}
\delta S = \frac{\delta M - \Omega \delta J}{T}.
\end{equation}
This relation gives two first derivatives of the entropy $S$ with respect to $M$ and $a$, respectively:
\begin{eqnarray}
&& \partial_M S = \frac{1 - \Omega a}{T} = \frac{2 \pi r_h^2}{r_h - M G - M r_h \partial_{r_h} G},
\label{eq_partialS1}
\\
&& \partial_a S = - \frac{\Omega M}{T} = - \frac{2 \pi M a}{r_h - M G - M r_h \partial_{r_h} G}.
\label{eq_partialS2}
\end{eqnarray}
By applying the relations~\eqref{eq_partialR}, we can rewrite $\partial_M S$ in terms of the area of the horizon~\eqref{area} as
\begin{equation}
\partial_M S = \frac{2 \pi r_h \partial_M r_h}{G + M \partial_M G} = \frac{\partial_M A_h}{4 (G + M \partial_M G)} .
\end{equation}
To ensure the consistency, i.e. $\partial_a \partial_M S = \partial_M \partial_a S$, which must be satisfied in order for~\eqref{cons} to give a well-defined entropy, a sufficient condition is that $G + M \partial_M G$ is a function of the area $A_h$ and then the derivative $\partial_a S$ can be expressed in the following form
\begin{equation}
\partial_a S = \frac{\partial_a A_h}{4 (G + M \partial_M G)}
= \frac{2 \pi (a + r_h \partial_a r_h)}{G + M \partial_M G}.
\end{equation}
By comparing this relation with~\eqref{eq_partialS2} and using~\eqref{eq_partialR}, we get the equation
\begin{equation} \label{eq_EqGh}
M a \partial_M G + r_h^2 \partial_a G - a r_h \partial_{r_h} G = 0,
\end{equation}
and its solution is
\begin{equation} \label{eq_solGh}
G(r_h, M, a) = G(M r_h, r_h^2 + a^2).
\end{equation}
We then see from~\eqref{hc} that $\Delta(r_h)$ reduces to just a function of $M r_h$ and $A_h$, and we can solve $\Delta(r_h) = 0$ for $M r_h$ as a function of $A_h$. This also means that $G + M \partial_M G$ becomes a function of area in Eq.~\eqref{eq_solGh}. It follows that the entropy is also a function of $A_h$ which can be computed by the general formula
\begin{equation} \label{entropy}
S = \int \frac{dA_h}{4 ( G + M \partial_M G)}.
\end{equation}
Here we have used a simple argument to derive the consistency equation~\eqref{eq_EqGh}. It can also be derived by a brute force from $\partial_a$ of~\eqref{eq_partialS1} equal to $\partial_M$ of~\eqref{eq_partialS2}. We emphasize that $M r_h$ and the area are related only on the horizon due to~\eqref{hc}, but they become independent when we extend away from the horizon because of the lack of~\eqref{hc}.

In summary, from the view points of the regularity of the horizon as well as the consistency with the zero-th and first laws of stationary black hole thermodynamics, it is necessary that the scale identification should be angle-independent and be in the form of Eq.~\eqref{eq_solGh} as a function of the mass and the area. Note that this result is quite general and valid beyond the asymptotic safety program; once the Newton coupling or equivalently the mass of the black holes are assumed to depend on the coordinates, we have to adopt Eq.~\eqref{eq_solGh}.

\section{Quantum Improved Regular Kerr Black Holes} \label{sec:4}

It is natural to assume that the result for running coupling~\eqref{eq_solGh} on the horizon may be extended away from the horizon, i.e. $G(r) = G(M r, r^2 + a^2)$. As a simple identification, we can consider
\begin{equation}
k^2 = \frac{f_1(A)}{(r^2 + a^2) (M r)^p} + f_2(A),
\end{equation}
where $f_1(A)$ and $f_2(A)$ are arbitrary functions of the area $A$ of suitable dimensions. The important point is that this diverges in the limit $r \to 0$, such that the Newton coupling~\eqref{Ncoupling} vanishes and the singularity is resolved.

As a simplest choice, let us take
\begin{equation} \label{choice2}
k^2 = \frac{\xi^2}{(r^2 + a^2) (M r)^p}.
\end{equation}
The quantum effect is then significant at
\begin{equation}
(r^2 + a^2) r^p \lesssim \frac{\tilde\omega}{M^p M_P^2}, \quad \Rightarrow \quad r^p \lesssim \frac{\tilde\omega}{a^2 M^p M_P^2}.
\end{equation}

This identification may appear to have a subtle property that the energy scale is divergent as the mass approaches to zero (flat spacetime)
\begin{equation}
\lim_{M \to 0} k^2 = \lim_{M \to 0} \frac{\xi^2}{(r^2 + a^2) (M r)^p} \to \infty.
\end{equation}
However this is just a manifestation of the expectation that the quantum effects are large for small black holes.

This turns out to give physically interesting results for the resolution of the singularity and other properties of the black holes. From Eq.~\eqref{Ncoupling}, we find the Newton coupling
\begin{equation} \label{eq_Id_p}
G(r) = G_0 \frac{(M r)^p (r^2 + a^2)}{(r^2 + a^2) (M r)^p + \tilde\omega G_0}.
\end{equation}
In the following, we are going to analyze properties, in particular the resolution of the ring singularity and the existence of CTCs, in the quantum improved Kerr black holes.

\subsection{Resolving Singularity} \label{sec:4A}

The singularities of a spacetime are characterised in the associated geometric invariants. In particular, for the rotating black holes the algebraically complete set consists of four invariants~\cite{Torres:2016pgk}, two of them, $R$ and $I_6$, are real and the other two, $I$ and $K$, are complex (their definitions are given in Appendix~\ref{sec:AA}). For the quantum improved Kerr black holes with running coupling~\eqref{eq_Id_p}, these four invariants take the form
\begin{eqnarray}
R &=& - \frac{2 \omega G_0^2 M^{p+1} r^{p-1} \Gamma_1}{(r^2 + a^2 \cos^2\theta) [(r^2 + a^2) (M r)^p + \omega G_0]^3},
\\
I_6 &=& \frac{\omega^2 G_0^4 M^{2p+2} r^{2p-2} \left[ \Gamma_1 (r^2 + a^2 \cos^2\theta) + \Gamma_2 \right]^2}{12 (r^2 + a^2 \cos^2\theta)^4 [(r^2 + a^2) (M r)^p + \omega G_0]^6},
\\
I &=& \frac{G_0^2 M^{2p+2} r^{2p-2} \left[ \omega G_0 \Gamma_1 (r + i a \cos\theta)^2 + \Gamma_3 (r + i a \cos\theta) - \Gamma_4 \right]^2}{36 (r^2 + a^2 \cos^2\theta)^2 (r + i a \cos\theta)^4 [(r^2 + a^2) (M r)^p + \omega G_0]^6},
\\
K &=& \frac{\omega^2 G_0^5 M^{3p+3} r^{3p-3}\left[ \Gamma_1 (r^2 + a^2 \cos^2\theta) + \Gamma_2 \right]^2 \left[ \omega G_0 \Gamma_1 (r + i a \cos\theta)^2 + \Gamma_3 (r + i a \cos\theta) - \Gamma_4 \right]}{12 (r^2 + a^2 \cos^2\theta)^5 (r + i a \cos\theta)^2 [(r^2 + a^2) (M r)^p + \omega G_0]^9},
\end{eqnarray}
where
\begin{eqnarray}
\Gamma_1 &=& (M r)^p \left[ (p + 1) (p + 2) r^4 + 2 (p^2 + p - 3) r^2 a^2 + p (p - 1) a^4 \right] - \omega G_0 \left[ (p + 2) (p + 3) r^2 + p (p + 1) a^2 \right],
\nonumber\\
\Gamma_2 &=& 4 r^2 \left[ 2 r^2 + p (r^2 + a^2) \right] \left[ (r^2 + a^2) (M r)^p + \omega G_0 \right],
\nonumber\\
\Gamma_3 &=& 6 r \left\{ M^{2 p} r^{2 p} (r^2 + a^2)^3 + \omega G_0 (M r)^p (r^2 + a^2) \left[ (p + 4) r^2 + (p + 2) a^2 \right] + \omega^2 G_0^2 \left[ (p + 3) r^2 + (p + 1) a^2 \right] \right\},
\nonumber\\
\Gamma_4 &=& 12 r^2 (r^2 + a^2) \left[ (r^2 + a^2) (M r)^p + \omega G_0 \right]^2.
\end{eqnarray}

Near the ``equatorial disk'', i.e. $r \to 0$, these quantities have the following fall off:
\begin{equation}
R \sim \left\{ \begin{array}{rl} r^{p - 1}, & \quad \theta \ne \pi/2 \\ r^{p - 3}, & \quad \theta = \pi/2 \end{array} \right., \qquad I_6 \sim I \sim \left\{ \begin{array}{rl} r^{2 (p - 1)}, & \quad \theta \ne \pi/2 \\ r^{2 (p - 3)}, & \quad \theta = \pi/2 \end{array} \right., \qquad K \sim \left\{ \begin{array}{rl} r^{3 (p - 1)}, & \quad \theta \ne \pi/2 \\ r^{3 (p - 3)}, & \quad \theta = \pi/2 \end{array} \right..
\end{equation}
Therefore, in order to avoid divergence in the limit $r \to 0$, the value of $p$ cannot be less than three. In the case $p = 3$, all invariants have nonvanishing values for $\theta = \pi/2$, thus there is a discontinuity at the edge of equatorial disk.\footnote{
This case is analogous to the rotating Hayward black holes by the change $M \to M r^3/(r^3 + \ell^3)$ in Kerr solution. However, the rotating Hayward black holes indeed do not have a consistent thermodynamics unless the replacement $\ell \to \ell/M$ is made.}
Moreover, if $p$ is an odd integer, the term $(M r)^p + \omega$ appearing in the denominator in all four invariants can vanish in the region $r < 0$ (both $M$ and $\omega$ are positive) and the curvature singularity occurs. To avoid the singularity, we consider the case that $p$ is an even integer larger than 3 (minimum 4).

\subsection{Closed Time-like Curves} \label{sec:4C}

As is well known, the Kerr black holes have CTCs which cause the violation of causality inside the inner Cauchy horizon. The metric component $g_{\varphi\varphi}$ ($G$ is a constant) is
\begin{equation}
g_{\varphi\varphi} = \sin^2\theta \left( r^2 + a^2 + \frac{2 G M r a^2 \sin^2\theta}{\Sigma} \right),
\end{equation}
which can be negative if $r < 0$ and $\Sigma$ is small enough.\footnote{
In Appendix~\ref{sec:4B}, we show that the geometric completeness requires the extension to negative $r$.}
For example, at $\theta = \pi/2$, it reduces to
\begin{equation} \label{ppc}
g_{\varphi\varphi}\Big|_{\theta=\pi/2} = r^2 + a^2 + \frac{2 G M a^2}{r} \quad
\stackrel{|r| \ll a}{\longrightarrow} \quad a^2 \left( 1 + \frac{2 G M}{r} \right) \quad
\stackrel{|r| \ll G M}{\longrightarrow} \quad \frac{2 G M a^2}{r} < 0, \quad \textrm{if} \quad r < 0,
\end{equation}
which could be negative for $r < 0$ and $r \ll G M, a$.

For the quantum improved Kerr black holes with the identification~\eqref{eq_Id_p} and even integer $p$ (to avoid curvature singularity in $r < 0$), we have
\begin{equation}
g_{\varphi\varphi} = \sin^2\theta \left( r^2 + a^2 + \frac{2 G_0 (M r)^{p+1} (r^2 + a^2) a^2 \sin^2\theta}{[(M r)^p (r^2 + a^2) + \tilde\omega G_0] \Sigma} \right),
\end{equation}
which again can be negative if $r < 0$ and $\Sigma$ is small enough. If $g_{\varphi\varphi}$ is positive, then there are no CTCs.

To get rough estimate if this becomes negative, let us take $\theta = \pi/2$
\begin{equation} \label{pp}
g_{\varphi\varphi}\Big|_{\theta=\pi/2} = (r^2 + a^2) \left( 1 + \frac{2 a^2 M^{p+1} r^{p-1}/\tilde\omega}{(r^2 + a^2) (M r)^p/(\tilde\omega G_0)+1} \right).
\end{equation}
The first factor is always positive, and the second term in the bracket is negatively decreasing\footnote{
Note that we are considering even $p$ in order to avoid the curvature singularity.}
until $(r^2 + a^2) (M r)^p/(\tilde\omega G_0)$ becomes of order one. This is roughly $r \sim -(\tilde\omega G_0 /a^2)^{1/p}/M$ for small $|r|$. Beyond this value, the negative term gets smaller in its absolute value, and would be no longer more negative because the power of $r$ is bigger in the denominator. At this point, we see that Eq.~\eqref{pp} is approximately
\begin{equation} \label{eq_gpp}
g_{\varphi\varphi}\Big|_{\theta=\pi/2} = \mbox{(positive factor)} \left( 1 - \frac{a^{2/p} M^2 G_0^{(p-1)/p}}{\tilde\omega^{1/p}} \right) .
\end{equation}
So $g_{\varphi\varphi}$ can be positive if this quantity is positive for small $|r|$.

It is clear that the CTC always exists in classical solution ($\tilde\omega = 0$). The sufficient condition for quantum effect to remove the CTC is
\begin{equation} \label{condi:CTCrmv}
1 < \dfrac{\tilde\omega^{1/p}}{G_0 M^2} , \qquad \dfrac{a^2}{G_0} < 1.
\end{equation}
Note that $\tilde\omega^{1/p}/G_0 M^2$ is a measure for the impact of the quantum gravitational effects on the geometry (see Eq.~(4.8) in Ref.~\cite{Bonanno:2000ep}).

\subsection{Black Holes Thermodynamics} \label{sec:4D}

In this subsection, we will summarize the thermodynamical properties for quantum improved black holes for the identification~\eqref{eq_Id_p}. The radius of horizon $r_h$ is determined by $\Delta(r_h) = 0$ which can be expressed as
\begin{equation}
r_h^2 + a^2 = G_0( 2 M r_h - \tilde\omega M^{-p} r_h^{-p} ).
\end{equation}
By introducing a new variable $\rho_h \equiv M r_h$ we can have following relations for horizon area $A_h = 4 \pi (r_h^2 + a^2)$, and its derivative
\begin{equation}
A_h = 4 \pi G_0 (2 \rho_h - \tilde\omega \rho_h^{-p}), \qquad d A_h = 4 \pi G_0 \rho_h^{-p-1} (2 \rho_h^{p+1} + p \tilde\omega) d\rho_h,
\end{equation}
and it is straightforward to compute
\begin{equation}
G + M \partial_M G = \frac{G_0 (r_h^2 + a^2) (M r_h)^p \left[ (r_h^2 + a^2) (M r_h)^p + (p + 1) \tilde\omega G_0 \right]}{\left[ (r_h^2 + a^2) (M r_h)^p + \tilde\omega G_0 \right]^2} = \frac{G_0 (2 \rho_h^{p+1} - \tilde\omega) \left( 2 \rho_h^{p+1} + p \tilde\omega \right)}{\left( 2 \rho_h^{p+1} \right)^2}.
\end{equation}
Therefore the entropy~\eqref{entropy} becomes
\begin{eqnarray}
S &=& 4 \pi \int \frac{\rho_h^{p+1}}{2 \rho_h^{p+1} - \tilde\omega} d\rho_h = 2 \pi \int \left( 1 - \frac{\tilde\omega}{2 \rho_h^{p+1}} \right)^{-1} d\rho_h = 2 \pi \int \left[ 1 + \frac{\tilde\omega}{2 \rho_h^{p+1}} + \left( \frac{\tilde\omega}{2 \rho_h^{p+1}} \right)^2 + \cdots \right] d\rho_h
\nonumber\\
&=& \frac{A_h}{4 G} - 2 \pi \sum_{j=1}^\infty \frac{\tilde\omega^j}{2^j (j p + j - 1)} \left( \frac{A_h}{8 \pi G} \right)^{-(j p + j -1)} = 2 \pi \rho_h  - \frac{\pi \tilde\omega}{p} \rho_h^{-p} F\left(1, \frac{p}{p+1}; \frac{2p+1}{p+1}; \frac{\tilde\omega}{2 \rho_h^{p+1}} \right).
\end{eqnarray}

The temperature for quantum improved black holes is
\begin{equation}
T = \frac{r_h - M G - M r_h \partial_{r_h} G}{2 \pi (r_h^2 +a^2)},
\end{equation}
and for the identification~\eqref{eq_Id_p}, it reduces to
\begin{equation}
T = \frac{r_h^2 - a^2}{4 \pi r_h (r_h^2 + a^2)} - \tilde\omega \frac{p G_0 (2 \rho_h^{p+1} - \tilde\omega) + 2 r_h^2 \rho_h^p}{8 \pi G_0 r_h \rho_h^{p+1} (2 \rho_h^{p+1} - \tilde\omega)}.
\end{equation}
Both quantum corrections for the entropy and temperature are negative.

\section{Conclusion} \label{sec:5}

In this paper we have studied the quantum improvement of Kerr black holes in asymptotically safe gravity. The main purpose is to construct regular black holes in which the ring singularity in classical Kerr solutions is resolved by quantum effect. Resolution of singularity indeed has been largely discussed in the literatures. But in those studies, the possible violation of the black hole thermodynamics has been simply neglected. As pointed out in the introduction, if the mass or equivalently the gravitational constant is a function of coordinate only, our results show that the thermodynamics are not satisfied. If we consider only the case that the running coupling does not depend on the mass parameter, we find the condition $G = G(r^2 + a^2)$~\cite{Chen:2022xjk} to be necessary for the first law of black hole thermodynamics. The phase structure in this case is analyzed~\cite{Chen:2023pcv}. Unfortunately, this type of Newton coupling does not approach to zero or diverge near the ring singularity and hence cannot resolve singularity.

What we have found in this paper is that when the mass parameter is included in the scale identification, there is more general identification $G = G(M r_h, A_h)$ that can preserve the consistency of black hole thermodynamics. Though $Mr_h$ is actually also a function of the area {\em on} the horizon, this form can be extended away from the horizon. In the extended region, $M r$ becomes an independent variable. We have then shown that this can resolve the singularity. Although we have studied only a particular ``solution'' for the Newton coupling which gives regular quantum-improved Kerr black holes, we expect that even for more general solutions for the Newton coupling, the qualitative feature would not change as far as the choice of the scale identification can resolve the singularities. The regularity of our black hole solutions is confirmed by four geometric scalar invariants (two real and 2 complex). All of them are continuous and never divergent in all ranges of radial coordinate, from negative infinity to positive infinity. We have also found that the geodesics are incomplete at $r = 0$ and therefore an extension to negative values of $r$ is necessary (Appendix~\ref{sec:4B}). Moreover, for the classical Kerr, CTCs always exist in $r < 0$.\footnote{
In $r < 0$ with sufficiently small $|r|$, $g_{\varphi\varphi}$ can be negative (it has a discontinuity at $r = 0$), as can be seen from Eq.~\eqref{ppc}, and therefore a closed orbit of the Killing vector field $\partial /\partial \varphi$ becomes a timelike orbit, hence a CTC in $r < 0$. By making a suitable deformation of such a closed timelike Killing orbit, one can construct a CTC which stretches any point inside $r_-$.}
It is remarkable that the occurrence of CTCs inside the Cauchy horizon is also eliminated by the quantum effects for the mass and angular momentum of black holes.
We have seen that for the quantum improved Kerr, there are no CTCs when $|r|$ is small enough, and $g_{\varphi\varphi}$ is smooth at $r = 0$. Therefore, CTCs do not exist for large $|r|$ and small $|r|$.

Although resolving classical singularities as well as CTCs is by itself a desirable property as a result of quantum gravity effects, our quantum improved Kerr metric admits a Cauchy horizon, and thus violates the strong cosmic censorship, as the other known quantum improved models also do~\cite{Bonanno:2000ep, Pawlowski:2018swz, Ishibashi:2021kmf}. Since a Cauchy horizon is, in general, the boundary of the maximal time evolution of given initial data, its existence jeopardizes the predictability of underlying theory. The Cauchy horizon inside the classical solutions such as the Reissner-Nordstrom and Kerr solutions is known to be unstable against a certain type of perturbations~\cite{Poisson:1989zz, Ori:1991zz}; such an instability is sometime referred to as the mass-inflation. It has recently been shown~\cite{Carballo-Rubio:2021bpr} that the essentially same type of Cauchy horizon instability also occurs in various regular (non-rotating) black hole models, including the quantum improved Schwarzschild black hole in the asymptotic safety scenario~\cite{Bonanno:2000ep}.
It is of considerable interest to study whether the Cauchy horizon instability occurs in our quantum improved Kerr geometry.
The unstable Cauchy horizon may possibly turn to be either a null or spacelike singularity so that the whole geometry recovers the global hyperbolicity. If it is indeed the case, the strong cosmic censorship is safe and the underlying gravitational theory can be deterministic.
If we demand the asymptotic safe quantum gravity to be truly a deterministic theory, then the appearance of Cauchy horizon should be prevented by some quantum effects or instabilities, resulting possibly in producing yet another type of singularities along (or even before the formation of) the would-be Cauchy horizon. How to reconcile simultaneously the resolution of singularities and the disappearance of Cauchy horizon is an important problem to gain a further understanding of possible consequences of quantum effects on spacetime geometries. It would be interesting to pursue the possibility of some scale identification which precludes the occurrence of Cauchy horizon from the first place.
In this respect, we note the possibility that a classically regular horizon can become a quantum singularity, or even before the formation of a classical Cauchy horizon, a quantum singularity appears~\cite{Bousso:2022tdb}. The idea of quantum singularity comes from a consideration of the semiclassical time evolution with the effective Newton constant dependent on some cutoff scale. It would be interesting to consider such a notion of quantum singularity in the context of the asymptotic safe quantum gravity.\footnote{
For this purpose, one should presumably have to introduce a suitable scale identification at the action level, derive the effective Einstein equations from the quantum improved action as a counterpart of the semiclassical Einstein equations, and examine the initial value problem. Since the effective Einstein equations derived from quantum improved action involve not only a scale-(position)-dependent Newton coupling but also its derivative, the resultant spacetime geometry could drastically be different from those obtained by quantum improvement at the solution level. This program is however beyond the scope of the present paper.}

\acknowledgments
We thank Aaron Held for valuable discussions.
The work of C.M.C. was supported by the National Science and Technology Council of the R.O.C. (Taiwan) under the grant 112-2112-M-008-020.
The work of A.I. was supported in part by JSPS KAKENHI Grants No. 21H05182, 21H05186, 20K03938, 20K03975, 17K05451, and 15K05092.
The work of N.O. was supported in part by the Grant-in-Aid for Scientific Research Fund of the JSPS (C) No. 20K03980 and Taiwan NSTC 112-2811-M-008-016.

\begin{appendix}

\section{Geometric Invariant Quantities} \label{sec:AA}
For the rotating black holes, the algebraically complete set of invariants are~\cite{Torres:2016pgk}
\begin{equation}
R, \qquad I_6 = \frac1{12} S_{\alpha\beta} S^{\alpha\beta}, \qquad I = \frac1{24} {\bar C}_{\alpha\beta\mu\nu} {\bar C}^{\alpha\beta\mu\nu}, \qquad K = \frac14 {\bar C}_{\alpha\mu\nu\beta} S^{\mu\nu} S^{\alpha\beta},
\end{equation}
where $S_{\mu\nu}$ and ${\bar C}_{\alpha\beta\mu\nu}$ are defined as
\begin{equation}
S_{\alpha\beta} = R_{\alpha\beta} - \frac14 g_{\alpha\beta} R, \qquad {\bar C}_{\alpha\beta\mu\nu} = \frac12 \left( C_{\alpha\beta\mu\nu} + i \, {}^* C_{\alpha\beta\mu\nu} \right), \quad {}^* C_{\alpha\beta\mu\nu} = \frac12 \epsilon_{\alpha\beta\gamma\delta} C^{\gamma\delta}{}_{\mu\nu},
\end{equation}
and $R, R_{\alpha\beta}$ and $C_{\alpha\beta\mu\nu}$ are the scalar curvature, Ricci tensor and Weyl tensor, respectively,
\begin{equation}
C_{\alpha\beta\mu\nu} = R_{\alpha\beta\mu\nu} + \frac12 \left( R_{\alpha\nu} g_{\beta\mu} - R_{\alpha\mu} g_{\beta\nu} + R_{\beta\mu} g_{\alpha\nu} - R_{\beta\nu} g_{\alpha\mu} \right) + \frac16 R \left( g_{\alpha\mu} g_{\beta\nu} - g_{\alpha\nu} g_{\beta\mu} \right).
\end{equation}
The Kretschmann scalar is related to these invariants as
\begin{equation}
\mathcal{K} = R_{\alpha\beta\mu\nu} R^{\alpha\beta\mu\nu} = C_{\alpha\beta\mu\nu} C^{\alpha\beta\mu\nu} + 2 R_{\mu\nu} R^{\mu\nu} - \frac13 R^2 = 48\, \mathrm{Re}(I) + 24 I_6 + \frac16 R^2.
\end{equation}

For the Kerr black holes in the BL coordinates~\eqref{eq_KerrBL} with the running coupling $G = G(r)$, these invariant quantities are
\begin{eqnarray} \label{eq_KerrR}
R &=& \frac{2 M (r \partial_r^2 G + 2 \partial_r G)}{r^2 + a^2 \cos^2\theta},
\\
I_6 &=& \frac{M^2 \left[ (r \partial_r^2 G + 2 \partial_r G) (r^2 + a^2 \cos^2\theta) - 4 r^2 \partial_r G \right]^2}{12 (r^2 + a^2 \cos^2\theta)^4},
\\
I &=& \frac{M^2 \left[ (r \partial_r^2 G + 2 \partial_r G) (r + i a \cos\theta)^2 - 6 (r \partial_r G + G) (r + i a \cos\theta) + 12 r G \right]^2}{36 (r^2 + a^2 \cos^2\theta)^2 (r + i a \cos\theta)^4},
\\
K &=& - \frac{M^3 \left[ (r \partial_r^2 G + 2 \partial_r G) (r^2 + a^2 \cos^2\theta) - 4 r^2 \partial_r G \right]^2}{12 (r^2 + a^2 \cos^2\theta)^5 (r + i a \cos\theta)^2}
\nonumber\\
&& \times \left[ (r \partial_r^2 G + 2 \partial_r G) (r + i a \cos\theta)^2 - 6 (r \partial_r G + G) (r + i a \cos\theta) + 12 r G \right], \label{eq_KerrK}
\end{eqnarray}
and the Kretschmann scalar is
\begin{eqnarray}
\mathcal{K} &=& \frac{4 M^2}{(r^2 + a^2 \cos^2\theta)^6} \Bigl\{ \Bigl[ (r \partial_r^2 G + 2 \partial_r G) (r^2 + a^2 \cos^2\theta)^2 - 2 r (2 r \partial_r G + 3 G) (r^2 + a^2 \cos^2\theta) + 8 r^3 G \Bigr]^2
\nonumber\\
&& - 12 (r \partial_r G + G)^2 (r^2 + a^2 \cos^2\theta)^3 + 4 r^2 (2 r \partial_r G + 3 G) (2 r \partial_r G + 15 G) (r^2 + a^2 \cos^2\theta)^2
\nonumber\\
&& - 32 r^4 G (4 r \partial_r G + 15 G) (r^2 + a^2 \cos^2\theta) + 320 r^6 G^2 \Bigr\}.
\end{eqnarray}

\section{Quantum Improved Kerr-(A)dS Black Holes} \label{sec:AB}

Here we generalize the analysis of thermodynamics in Sec.~\ref{sec:3} to the case with a fixed cosmological constant $\Lambda$. This case may not be directly related to the asymptotically safe gravity (see, however Ref.~\cite{KO}), but it is still interesting to check the constraint on the running coupling with the assumption $G = G(r, M, a)$. Here the cosmological constant is supposed to be constant (not variable) in the first law of thermodynamics. The line-element of the classical Kerr-(A)dS black hole solution in the BL coordinate is
\begin{equation}
ds^2 = - \frac{\Delta_r}{\Sigma} \left( dt - \frac{a \sin^2\theta}{\Xi} d\varphi \right)^2 + \frac{\Sigma}{\Delta_r} dr^2 + \frac{\Sigma}{\Delta_\theta} d\theta^2 + \frac{\Delta_\theta}{\Sigma} \sin^2\theta \left( a dt - \frac{r^2 + a^2}{\Xi} d\varphi \right)^2,
\end{equation}
where
\begin{equation} \label{eq_KerrAdS}
\Delta_r = (r^2 + a^2) \left( 1 - \frac{\Lambda}3 r^2 \right) - 2 G M r, \quad
\Delta_\theta = 1 + \frac{\Lambda}3 a^2 \cos^2\theta, \quad
\Sigma = r^2 + a^2 \cos^2\theta, \quad \Xi = 1 + \frac{\Lambda}3 a^2.
\end{equation}
In the semiclassical case, where $G$ is a constant, the Hawking temperature and the entropy are given by
\begin{equation} \label{eq_TSAdS}
T = \frac{\kappa}{2\pi} = \frac{r_h}{4 \pi (r_h^2 + a^2)} \left( 1 - \frac{a^2}{r_h^2} - \frac{\Lambda a^2}3 - \Lambda r_h^2 \right), \qquad
S = \frac{A_h}{4 G},
\end{equation}
where $A_h$ is the horizon area
\begin{equation} \label{areads}
A_h = \frac{4\pi (r_h^2 + a^2)}{\Xi}.
\end{equation}
These quantities satisfy the first law of thermodynamics
\begin{equation} \label{eq_firstlawAdS}
dE = T dS + \Omega dJ,
\end{equation}
where
\begin{equation} \label{eq_EJOAdS}
E = \frac{M}{\Xi^2}, \qquad J = \frac{M a}{\Xi^2}, \qquad \Omega = \frac{a (1 - \Lambda r_h^2/3)}{r_h^2 + a^2}.
\end{equation}

Note again that these formulas hold not only for the outer event horizon, but also for the inner Cauchy horizon, as well as for cosmological horizon when $\Lambda > 0$.

The Kretschmann scalar is
\begin{equation} \label{KerrAdSKretschmann}
\mathcal{K} = \frac83 \Lambda^2 + \frac{48 M^2 G^2 (r^6 - 15 a^2 \cos^2\theta r^4 + 15 a^4 \cos^4\theta r^2 - a^6 \cos^6\theta)}{(r^2 + a^2 \cos^2\theta)^6},
\end{equation}
which depends on the angle coordinate $\theta$. Taking the limit $r \to 0$ we have
\begin{equation}
\lim_{r \to 0} \mathcal{K} =
\begin{cases}
\dfrac{48 M^2 G^2}{r^6} \to \infty, &
\theta = \pi/2 \\
&
\\
\frac83 \Lambda^2 - \dfrac{48 M^2 G^2}{a^6 \cos^6\theta}, & \theta \ne \pi/2
\end{cases}.
\end{equation}
This indicates a ring singularity at $r = 0, \, \theta = \pi/2, \varphi = \mbox{arbitrary}$, and the Kretschmann scalar is not a continuous function in $\theta$ on the disk $r = 0$ at the ring singularity.

For the quantum improved rotating black holes with a cosmological constant, the horizon radius is determined by
\begin{equation} \label{eq_DrhL}
\Delta_r(r_h) = (r_h^2 + a^2) \left( 1 - \frac{\Lambda}3 r_h^2 \right) - 2 G(r_h; M, a, \Lambda) M r_h = 0,
\end{equation}
and its first derivative, treating $r_h = r_h(M, a)$, gives
\begin{eqnarray}
0 = d\Delta_r(r_h) &=& 2 r_h (\partial_M r_h \, dM + \partial_a r_h \, da) + 2 a \, da - \frac{2 \Lambda}3 r_h (2 r_h^2 + a^2) (\partial_M r_h \, dM + \partial_a r_h \, da) - \frac{2 \Lambda}3 r_h^2 a \, da
\nonumber\\
&-& 2 r_h G \, dM - 2 M r_h \partial_M G \, dM - 2 M r_h \partial_a G \, da - 2 M (G + r_h \partial_{r_h} G) (\partial_M r_h \, dM + \partial_a r_h \, da).
\end{eqnarray}
This gives
\begin{eqnarray} \label{eq_partialRL}
&& \partial_M r_h = \frac{r_h (G + M \partial_M G)}{r_h - M G - M r_h \partial_{r_h} G - \frac{\Lambda}3 r_h (2 r_h^2 + a^2)},
\nonumber\\
&& \partial_a r_h = - \frac{a - M r_h \partial_a G - \Lambda r_h^2 a/3}{r_h - M G - M r_h \partial_{r_h} G - \frac{\Lambda}3 r_h (2 r_h^2 + a^2)}.
\end{eqnarray}
For Kerr-(A)dS black holes, the internal energy, angular momentum, angular velocity are given in~\eqref{eq_EJOAdS} and the temperature is
\begin{equation} \label{dstemp}
T = \frac{r_h - M G - M r_h \partial_{r_h} G - \frac{\Lambda}3 r_h (2 r_h^2 + a^2)}{2 \pi (r_h^2 + a^2)} = \frac{r_h (G + M \partial_M G)}{2 \pi (r_h^2 + a^2) \partial_M r_h},
\end{equation}
where we  have used \eqref{eq_partialRL} in the second equality. The thermodynamical first law $\delta E = T \delta S + \Omega \delta J$ gives
\begin{equation}
\delta S = \frac{\delta E - \Omega \delta J}{T}
\end{equation}
leading to
\begin{equation}
\partial_M S = \frac{r_h^2}{T \Xi (r_h^2 + a^2)}, \qquad \partial_a S = - \frac{M a (1 + \Lambda r_h^2)}{T \Xi^2 (r_h^2 + a^2)}.
\end{equation}
Using~\eqref{dstemp}, we find
\begin{equation} \label{eq_partialSL}
\partial_M S = \frac{2 \pi r_h \partial_M r_h}{\Xi (G + M \partial_M G)}, \qquad
\partial_a S = - \frac{2 \pi M a (1 + \Lambda r_h^2)\partial_M r_h}{\Xi^2 r_h(G+M\partial_M G)}.
\end{equation}
Again we can rewrite $\partial_M S$ in terms of horizon area~\eqref{areads}:
\begin{equation}
\partial_M S = \frac{\partial_M A_h}{4 (G + M \partial_M G)}.
\end{equation}
The consistency is ensured if $\partial_a S$ has the following form
\begin{equation}
\partial_a S = \frac{\partial_a A_h}{4 (G + M \partial_M G)} = \frac{2 \pi}{G + M \partial_M G} \left( \frac{r_h \partial_a r_h}{\Xi} + \frac{a (1 - \Lambda r_h^2/3)}{\Xi^2} \right).
\end{equation}
Comparing this equation with~\eqref{eq_partialSL} and using~\eqref{eq_partialRL} and~\eqref{eq_DrhL}, we can get the following equation for a consistent mass-dependent running coupling:
\begin{equation}
M a (1 + \Lambda r_h^2) \partial_M G(r_h, M, a) + r_h^2 \Xi \partial_a G(r_h, M, a) - a r_h \left( 1 - \frac{\Lambda}3 r_h^2 \right) \partial_{r_h} G(r_h, M, a) = 0.
\end{equation}
The solution of this equation is
\begin{equation}
G(r_h, M, a) = G\left( \frac{M r_h}{\Xi^2}, \frac{r_h^2+a^2}{\Xi} \right).
\end{equation}
For this case, it is not obvious that this solution is the function of the horizon area. To confirm this relation, we first divide equation~\eqref{eq_DrhL} by $\Xi^2$:
\begin{equation}
\frac{\Delta(r_h)}{\Xi^2} = \left( \frac{r_h^2 + a^2}{\Xi} \right) \left( \frac{1 - \frac{\Lambda}3 r_h^2}{\Xi} \right) - 2\, G\left( \frac{M r_h}{\Xi^2}, \frac{r_h^2 + a^2}{\Xi} \right) \frac{M r_h}{\Xi^2} = 0,
\end{equation}
where
\begin{equation}
\frac{1 - \frac{\Lambda}3 r_h^2}{\Xi} = 1 - \frac{\Lambda}3 \frac{r_h^2 + a^2}{\Xi} = 1 - \frac{\Lambda}{12 \pi} A_h.
\end{equation}
This is nothing but a function of $A_h$. Then we can easily see $M r_h/\Xi^2$ is also a function of $A_h$. The entropy for this system is again given by the general formula~\eqref{entropy}.

\section{On analytic extension and geodesic incompleteness} \label{sec:4B}


We have seen in Sec.~\ref{sec:4} that under our proposed scale identification~(\ref{choice2}) and the condition~(\ref{condi:CTCrmv}), the quantum effects can remove both the ring singularity and CTCs located inside the Cauchy horizon of the classical Kerr metric. As is well known, the interior region $-\infty < r < r_- := G_0 M - \sqrt{(G_0 M)^2 - a^2}$ of the classical Kerr geometry can only be obtained by the maximal {\em analytic} extension of the Kerr metric from positive value $r_-$ to large negative values $r \rightarrow - \infty$ (except the ring singularity at $r = 0$ and $\theta = \pi/2$). It is therefore natural to ask whether or not our quantum improved Kerr geometry is also analytic. Here, we address this issue by studying the behavior of null geodesics in our quantum improved Kerr black holes. We basically follow general analysis of the analytic continuation of causal geodesics made for some known regular black holes~\cite{Zhou:2022yio, Zhou:2023lwc}.

For massless particles, their trajectory $\{ t(\lambda), r(\lambda), \theta(\lambda), \varphi(\lambda) \}$ satisfy the constraint $ds^2 = 0$. In stationary, axisymmetric black holes, there are two conserved quantities associated with the Killing vectors $\xi^\mu = (1, 0, 0, 0)$ and $\eta^\mu = (0, 0, 0, 1)$. These are energy and angular momentum:
\begin{equation} \label{eq_def_el}
e = - \xi^\mu u^\nu g_{\mu\nu} = - g_{tt} \dot t - g_{t\varphi} \dot\varphi, \qquad
l = \eta^\mu u^\nu g_{\mu\nu} = g_{\varphi t} \dot t + g_{\varphi\varphi} \dot\varphi.
\end{equation}
Here {\it dot} denotes the derivative with respect to the affine parameter $\lambda$.

Let us consider the simplest case when a massless probe particle moves along the rotation axis, i.e. $\theta = \dot\theta = 0$. This implies $g_{t\varphi} = g_{\varphi\varphi} = 0$ and $l = 0$. For the quantum improved Kerr metric, the constraint and converged quantities reduce to
\begin{equation}
\left( 1 - \frac{2 G(r) M r}{r^2 + a^2} \right) \dot t^2 - \left( 1 - \frac{2 G(r) M r}{r^2 + a^2} \right)^{-1} \dot r^2 = 0, \qquad
e = \left( 1 - \frac{2 G(r) M r}{r^2 + a^2} \right) \dot t, \qquad l = 0,
\end{equation}
which lead to ($t' \equiv d t/d r$)
\begin{equation}
\dot r^2 = e = \text{constant}, \qquad t'^2 = (\dot t/\dot r)^2 = \left( 1 - \frac{2 G(r) M r}{r^2 + a^2} \right)^{-2}.
\end{equation}

The behavior of $t'$ in the $r \to 0$ is essential. Let us check this behavior for two well-known ``regular'' black holes. For the rotating Hayward black holes, we can view effectively $G(r) = G_0 r^3/(r^3 + \ell^3)$ and find
\begin{eqnarray}
t_H'^2 &=& 1 + \frac{4 G_0 M}{\ell^3} \left( \frac{r^4}{a^2} - \frac{r^6}{a^4} + \frac{r^8}{a^6}
- \frac{r^{10}}{a^8} + \cdots \right) - \frac{4 G_0 M}{\ell^6} \left( \frac{r^7}{a^2} - \frac{r^9}{a^4}
+ \frac{r^{11}}{a^6} - \cdots \right) + \frac{4 G_0 M}{\ell^9} \left( \frac{r^{10}}{a^2} - \cdots \right) + \cdots
\nonumber\\
&& \quad + \frac{12 G_0^2 M^2}{\ell^6} \left( \frac{r^8}{a^4} - 2 \frac{r^{10}}{a^6}
+ 3 \frac{r^{12}}{a^8} - \cdots \right) - \frac{24 G_0^2 M^2}{\ell^9} \left( \frac{r^{11}}{a^4}
- 2 \frac{r^{13}}{a^6} + \cdots \right) + \cdots .
\end{eqnarray}
Similarly, for the Bardeen black holes, we can consider $G(r) = G_0 r^3/(r^2 + \ell^2)^{3/2}$  effectively and obtain
\begin{eqnarray}
t_B'^2 &=& 1 + \frac{4 G_0 M}{\ell^3} \left( \frac{r^4}{a^2} - \frac{r^6}{a^4} + \frac{r^8}{a^6}
- \frac{r^{10}}{a^8} + \cdots \right) - \frac{6 G_0 M}{\ell^5} \left( \frac{r^6}{a^2} - \frac{r^8}{a^4}
+ \frac{r^{10}}{a^6} - \cdots \right) + \frac{15 G_0 M}{2 \ell^7} \left( \frac{r^8}{a^2} - \cdots \right) + \cdots
\nonumber\\
&& \quad + \frac{12 G_0^2 M^2}{\ell^6} \left( \frac{r^8}{a^4} - 2 \frac{r^{10}}{a^6} + 3 \frac{r^{12}}{a^8}
- \cdots \right) - \frac{36 G_0^2 M^2}{\ell^8} \left( \frac{r^{10}}{a^4} - 2 \frac{r^{12}}{a^6} + \cdots \right)
+ \cdots .
\end{eqnarray}
These two types of black holes have very different behaviors. For the rotating Hayward black holes, it was argued in~\cite{Zhou:2023lwc} that the $r^7$ (odd power) term flips the sign and this implies the geodesic is not analytic at $r = 0$. In this case, if one requires the analyticity in the unique continuation of geodesics, one should extend the geodesic to negative value of $r$. On the contrary, for the Bardeen black holes, there are only even powers of $r$, and this implies that the geodesic can return back via the corresponding antipodean point. Therefore, it is unnecessary to have an extension to negative value of $r$.

For the quantum improved Kerr black holes with $G(r) = G_0 (M r)^p (r^2 + a^2)/[(M r)^p (r^2 + a^2) + \tilde\omega G_0]$ we have
\begin{eqnarray}
t'^2 &=& 1 + 4 \frac{M^{p+1} r^{p+1}}{\tilde\omega} - 4 \frac{M^{2p+1} r^{2p+1} (r^2 + a^2)}{\tilde\omega^2 G_0} + 12 \frac{M^{2p+2} r^{2p+2}}{\tilde\omega^2}
\nonumber\\
&& + \, 4 \frac{M^{3p+1} r^{3p+1} (r^2 + a^2)^2}{\tilde\omega^3 G_0^2} - 24 \frac{M^{3p+2} r^{3p+2} (r^2 + a^2)}{\tilde\omega^3 G_0} + 32 \frac{M^{3p+3} r^{3p+3}}{\tilde\omega^3}
\\
&& - \, 4 \frac{M^{4p+1} r^{4p+1} (r^2 + a^2)^3}{\tilde\omega^4 G_0^3} + 36 \frac{M^{4p+2} r^{4p+2} (r^2 + a^2)^2}{\tilde\omega^4 G_0^2} - 96 \frac{M^{4p+3} r^{4p+3} (r^2 + a^2)}{\tilde\omega^4 G_0} + 80 \frac{M^{4p+4} r^{4p+4}}{\tilde\omega^4} + \cdots.
\nonumber
\end{eqnarray}
It is obvious that odd powers of $r$, for example $r^{2p+1}$, always exist. Similarly to the rotating Hayward black holes, the geodesics are not smooth at $r = 0$ for any value of $p$. The extension to negative value of $r$ is inevitable, as far as the unique continuation is required.

\end{appendix}


\end{document}